\documentclass[12pt,letterpaper]{article}
\usepackage[includeheadfoot, 
            marginratio={1:1,2:3}, 
            width=412pt, 
            height=688pt,]{geometry}

\usepackage{amsmath}
\usepackage{amsfonts}
\usepackage{amssymb}
\usepackage{bbm}

\usepackage{graphicx}

\allowdisplaybreaks[2]
\numberwithin{equation}{section}

\newcommand{\thba}[2]{[\!\!\begin{array}{c}
  {\phantom{}\vspace{-.5mm}\scriptstyle#1}
   \\[-1.6mm]{\scriptstyle #2}\end{array}\!\!]}

\newcommand{\be}{\begin{equation}}
\newcommand{\ee}{\end{equation}}
\newcommand{\bea}{\begin{eqnarray}}
\newcommand{\eea}{\end{eqnarray}}
\newcommand{\mbb}{\mathbb}

\def\IR{\relax{\rm I\kern-.18em R}}

\def\IP{\relax{\rm I\kern-.18em P}}
\def\inbar{\vrule height1.5ex width.4pt depth0pt}
\def\IC{\relax\,\hbox{$\inbar\kern-.3em{\rm C}$}}

\def\K3{{\bf K3}}

\DeclareMathOperator{\sign}{sign}

\begin{document}


\vspace*{-1.5cm}
\begin{flushright}
  {\small
  MPP-2007-55 \\
  LMU-ASC 30/07 \\
  }
\end{flushright}

\vspace{1.5cm}
\begin{center}
  {\LARGE
Thresholds for Intersecting D-branes Revisited\\
  }
\end{center}

\vspace{0.25cm}
\begin{center}
  {\small
  Nikolas Akerblom$^1$, Ralph~Blumenhagen$^1$, Dieter L\"ust$^{1,2}$, \\
  Maximilian Schmidt-Sommerfeld$^1$ \\
  }
\end{center}

\vspace{0.1cm}
\begin{center}
  \emph{$^{1}$
  Max-Planck-Institut f\"ur Physik, F\"ohringer Ring 6, \\
  80805 M\"unchen, Germany } \\
  \vspace{0.25cm}
  \emph{$^{2}$ Arnold-Sommerfeld-Center for Theoretical Physics, \\
  Department f\"ur Physik, Ludwig-Maximilians-Universit\"at  M\"unchen, \\
  Theresienstra\ss e 37, 80333 M\"unchen, Germany} \\
\end{center}

\vspace{-0.1cm}
\begin{center}
  \tt{
  akerblom, blumenha, luest, pumuckl@mppmu.mpg.de \\
  }
\end{center}

\vspace{1.5cm}
\begin{abstract}
\noindent 
Gauge threshold corrections for intersecting
D6-brane string models on toroidal orbifold
backgrounds are reconsidered.
Both by dimensionally regularising the appearing
open string one-loop diagrams in tree-channel as well as by zeta-function
regularisation of the corresponding loop-channel
one-loop diagrams, we arrive at a result 
which takes into account the infrared divergence from the contribution of the massless states in the running
of the gauge coupling constant as well as the contribution of states, which become light
in certain regions of the moduli space.

\end{abstract}

\thispagestyle{empty}
\clearpage
\tableofcontents


\section{Introduction}
Models based on orientifolds of type IIA/B string theory \cite{Uranga:2003pz,Lust:2004ks,Blumenhagen:2005mu,Blumenhagen:2006ci} have become an alternative to heterotic constructions in studying low energy effects of string theory. In contrast to the latter, different gauge groups are usually localised on different brane stacks which implies that the tree level gauge couplings vary from stack to stack. For a high string scale, this can be at variance with gauge coupling unification as it appears in the MSSM. Therefore, corrections to the gauge coupling constants in intersecting D6-brane models are quite important if one wants to build semi-realistic models and eventually make contact with experiment.

Recently, it was found that the very same quantities also
appear in the computation of D-instanton corrections
to such intersecting D6-brane models. In this context they quantify
the one-loop determinants of the fluctuations around
the E2-instanton \cite{Blumenhagen:2006xt,Abel:2006yk,Akerblom:2006hx}\footnote{For related  recent work on D-instanton effects see \cite{Billo:2002hm,Ibanez:2006da,Bianchi:2007fx,Cvetic:2007ku,Argurio:2007vq,Bianchi:2007wy,Ibanez:2007rs}.}.

For intersecting D6-branes on toroidal backgrounds, 
these threshold corrections
have been computed explicitly in \cite{Lust:2003ky}. (For a calculation in
type I models see \cite{Bachas:1996zt,Antoniadis:1999ge,Berg:2004ek,Bianchi:2005sa}.)
These results were generalised to Gepner models in \cite{Anastasopoulos:2006hn}.
In this paper we would like to revisit the actual computation
performed in \cite{Lust:2003ky}. Special care has to be taken of the
two different divergences appearing in the
relevant annulus and M\"obius diagrams. Namely, there
are infrared divergences stemming from
massless open string modes as well as ultraviolet divergences,
which are due to massless closed string tadpoles and which
sum to zero upon invoking the tadpole cancellation condition. 
We use two different
regularisation methods. First, we compute in tree channel, where
the divergence due to the tadpole can be extracted explicitly.
The infrared divergence is then taken care of by dimensional
regularisation.
Second, we perform the computation entirely in loop channel.
Here, the infrared divergence is manifest and can be subtracted
explicitly and the ultraviolet divergence is dealt with using
zeta-function regularisation of
divergent series.
Both methods give the same result in sectors preserving
${\cal N}=1$ supersymmetry\footnote{This is actually also true for
${\cal N}=2$ sectors but, as the results agree with \cite{Lust:2003ky},
the derivation will not be displayed here.}. However, this result differs slightly from
the one given in \cite{Lust:2003ky}.
The aim of this letter is to clarify this subtle issue.

In section 2 we display the method used to derive the one loop
corrections to the gauge couplings. In section 3 we perform the actual
calculations (in tree channel) and in section 4 we discuss the results and their relation
to \cite{Lust:2003ky}. The loop channel calculation is sketched in the appendix.


\section{One-loop thresholds for intersecting D6-branes on $\mbb T^6$}

The one-loop corrections to the gauge coupling constants can be computed
by means of the background field method, which essentially amounts to
computing the partition function in the presence of a magnetic field in
the four-dimensional space-time.

The gauge coupling constants of the various gauge group factors $G_a$,
up to one loop, have the following form
\bea
\label{thres}
 \frac{1}{g_a^2(\mu)} = \frac{1}{g_{a,\mathrm{string}}^2} +
                     \frac{b_a}{16 \pi^2}\, \ln\left( \frac{M_s^2}{\mu^2}\right) + \Delta_a \,, \label{gceffST} 
\eea
where $b_a$ is the beta function coefficient.
The first
term corresponds to the gauge coupling constant at the string scale, which contains
the tree-level gauge coupling as well as universal contributions
at one-loop,  the second term gives the usual one-loop
running of the coupling constants, and the third term denotes the one-loop string threshold
corrections originating from integrating
out massive string excitations.
All terms are encoded in the
aforementioned partition functions and can therefore be determined by calculating
all annulus and M\"obius diagrams with at least one boundary on the brane
where the gauge group factor $G_a$ is localised. As we will discuss at the end of this
letter, there is a subtle issue concerning the contribution
of massive states in $\Delta_a$, which become lighter than
the string scale $M_s$
for small intersection angles.

For the contribution of an annulus diagram to the threshold corrections 
for relatively supersymmetric intersecting branes,
the background field method gives the general expression
\bea
\label{threshold}
  T^A({\rm D6}_a,{\rm D6}_b)=\int_0^\infty  \frac{dt}{t}\, 
  \sum_{\alpha,\beta\neq(\frac{1}{2},\frac{1}{2})} (-1)^{2(\alpha+\beta)}\, 
 \frac{\vartheta''\thba{\alpha}{\beta}(it)}{\eta^3(it)} \,\, 
  A^{\rm CY}_{ab}\thba{\alpha}{\beta}(it) \;,
\eea
where  $A^{\rm CY}_{ab}$ denotes the annulus partition function
in the $(ab)$ open string sector of the internal ${\cal N}=2$
superconformal field theory describing the Calabi-Yau manifold. 
So far, the gauge thresholds can only be explicitly computed for toroidal orbifold
or Gepner models \cite{Anastasopoulos:2006hn}.

Let us from now on specialise to the case of the toroidal 
$\mbb Z_2 \times \mbb Z_2$
orbifold. In general, besides the O6-planes, an intersecting
D-brane model contains various stacks of D6-branes
wrapping factorisable supersymmetric three-cycles
defined by three pairs of  wrapping numbers $(m_a^I,n_a^I)$, $I=1,2,3$.
On each D6-brane we assume a gauge symmetry $U(N_a)$ and
we are interested in the one-loop gauge threshold corrections
to the gauge couplings of these $U(N_a)$ gauge symmetries. 
These one-loop thresholds are given by annulus and M\"obius
diagrams with one boundary on the $U(N_a)$ brane.


\section{Thresholds for ${\cal N}=1$ sectors}\label{thresholds}

We now come to the actual regularisation of amplitudes. We take as our starting point the raw amplitudes found in \cite{Lust:2003ky}. For the annulus diagram in an ${\cal N}=1$ sector, the expression to be examined is
\bea
 T^A({\rm D6}_a,{\rm D6}_b)&=&
  \frac{i I_{ab} N_b}{2 \pi} \int_0^\infty \frac{dt}{t}\,\, \sum_{I=1}^3\,\,
         \frac{\vartheta_1'}{\vartheta_1}
      {\textstyle \left(\frac{i\theta_{ab}^I t}{2},\frac{it}{2}\right) }\nonumber\\
 &=& -\frac{ I_{ab} N_b}{\pi} \int_0^\infty dl\,\, \sum_{I=1}^3\,\,
         \frac{\vartheta_1'}{\vartheta_1} {\textstyle (-\theta_{ab}^I,2il)}\, ,
 \label{annu4ab}
\eea
where $I_{ab}$ is the intersection number, $N_b$ is the number of branes on stack $b$ and $l=1/t$.
Additionally, $\pi \theta^I_{ab}$ is the intersection angle
of branes $a$ and $b$ on the $I$'th torus. Supersymmetry then imposes the `angle condition'
\begin{equation}\label{angle}
\sum_{I=1}^3 \theta_{ab}^I = 0\, ,
\end{equation}
which we assume to be fulfilled.

As it stands, \eqref{annu4ab} is divergent. As mentioned in the introduction,
there are two sources for this divergence: In the $q$-series of the $l$-channel integrand there is a constant (i.e. $q^0$-) term, giving us something proportional to $\int\! dl$. Even if this is subtracted (or thought of as taken care of by tadpole cancellation) there remains a divergence from $l\rightarrow 0$. The latter is
the same as the divergence for $t\rightarrow\infty$ coming from the constant term in the loop channel. It is therefore seen to be a logarithmic divergence which arises from the massless open string states and encodes the one-loop
running of the gauge couplings. Therefore, it will later on be replaced by $\ln\left( \frac{M_s^2}{\mu^2}\right)$.

Since, following \cite{Lust:2003ky}, we impose the tadpole cancellation condition, we do not worry about the $q^0$-divergence, but clearly something needs to be done about the remaining logarithmic divergence. There are at least two ways to proceed. One is to subtract this divergence in the $t$-channel and another is to employ `dimensional regularisation'. If one wants to extract the tadpole divergence manifestly, which is only possible in tree channel, at least the large $l$ (i.e. small $t$) part of the amplitude has to be calculated in tree channel. This means that one can only work in the large $t$ regime of the loop channel or, in other words, the $t$-integration has to be cut off at a finite lower limit. 
Since the computation is then difficult to do analytically, we shall carry out the tree channel computation using `dimensional regularisation'\footnote{This method of regularisation has already been put to work in \cite{Lust:2003ky}, however with a slightly different result than ours. We will discuss the difference
between this result and the result of \cite{Lust:2003ky} in the final section of the paper.
\label{fuss}}.

By carrying out the instructions given just before \eqref{appformula}\footnote{Note the seemingly different formula for $\frac{\vartheta'_1}{\vartheta_1}$ as the one in \cite{Lust:2003ky}. The latter can be brought into the form displayed here by performing the sum over $k$ there.} it is easy to derive the following formula:
\begin{equation}\label{tad}
- \frac{1}{\pi} \int_0^\infty\! dl\, \frac{\vartheta'_1}{\vartheta_1}(\theta,2il)
= -\cot(\pi \theta) \int_0^\infty\! dl + A\, ,
\end{equation}
where
\begin{equation}
A=4 i \int_0^\infty \!dl\!\sum_{m,n=1}^\infty \exp(-4 \pi l m n)\, \sinh(2 \pi i \theta m)\,. \label{doublesumint}
\end{equation}
When taking the sum over the $\theta$s, \eqref{tad} gives us \eqref{annu4ab} (up to prefactors), so we might as well regularise \eqref{tad}.
The first term in \eqref{tad} is the tadpole and the second needs to be regularised. To this end, we let $\int\!dl \to \int\! dl\,l^\epsilon$, where $\epsilon$ is a (small) positive number. In loop channel this amounts to $\int\!\frac{dt}{t} \to \int\! \frac{dt}{\,t^{1+\epsilon}}\simeq \frac{1}{\epsilon}$.
In analogy to the heterotic string \cite{Kaplunovsky:1992vs}, it is therefore justified to later substitute $\ln\left( \frac{M_s^2}{\mu^2}\right)$ for $\frac{1}{\epsilon}$. Integrating over $l$ and carrying out the sum over $n$ in
\eqref{doublesumint} yields
\begin{equation}
A=- \frac{1}{\pi} \sum_{m=1}^\infty \frac{\sin(2 \pi \theta m)}{m(4\pi m)^\epsilon} \,\,
  \Gamma(1+\epsilon)\, \zeta(1+\epsilon)\,.
\end{equation}
For $\epsilon\ll 1$, we can expand
\begin{equation}\label{aexpr}
A=-\frac{1}{\epsilon}\left(\frac{1}{\pi} \sum_{m=1}^\infty \frac{\sin(2 \pi \theta m)}{m}\right)
  + \frac{1}{\pi} \sum_{m=1}^\infty \frac{\ln(4 \pi m)}{m} \, \sin(2 \pi \theta m)+O(\epsilon)\, ,
\end{equation}
from which the $\epsilon \to 0^+$ divergence is nicely read off. The term in parentheses is a standard example of a Fourier series. In the open interval $(0,1)$ it sums to $1/2-\theta$, while for $\theta=0$ it gives zero and elsewhere it sums to formulas given by `periodic continuation' with period $1$. Upon performing the sum over $I$ in \eqref{annu4ab} and using \eqref{angle} one finds that the $\frac{1}{\epsilon}$-term is multiplied by a constant, which, taking into account the prefactors in \eqref{annu4ab}, is the contribution of brane stack $b$ to the beta function. Thus,
upon $\frac{1}{\epsilon} \to \ln\left( \frac{M_s^2}{\mu^2}\right)$, the correct one-loop running is reproduced.
 It remains to sum the left-over infinite series\footnote{By Dirichlet's test, it converges for all $\theta$ in,
 say, the open interval $(0,1)$. Moreover, it converges trivially to zero for $\theta=0$, and therefore it
 converges for all $\theta$ by periodicity.} in \eqref{aexpr}.

It turns out that, for $0<\theta<1$,
\begin{equation}\label{fourier}
\frac{1}{\pi} \sum_{m=1}^\infty \frac{\ln(4 \pi m)}{m}\, \sin(2 \pi \theta m)=\frac{1}{2}\ln\left(\frac{\Gamma(\theta)}{\Gamma(1-\theta)}\right)-\left(\ln 2-\gamma\right)\left(\theta-1/2\right)\, ,
\end{equation}
where $\gamma$ is the Euler--Mascheroni constant.

A way to derive the expression on the right hand side of this equation is presented in 
appendix \ref{app}. Presently, let us verify  that the relation (\ref{fourier}) is true. 
The idea is to interpret the left hand side  of \eqref{fourier} as the Fourier series of its right
hand side  (for the theory of Fourier series of functions with infinities like the one at hand see e.g. \cite{Hobson}).

The even terms in this Fourier series are all zero for reasons of symmetry\footnote{The rhs of \eqref{fourier} is odd under reflection of $\theta$ in $1/2$, i.e. under $\theta\to 1-\theta$.}, while for the odd terms we have to calculate the sine Fourier coefficients
\begin{equation}
b_m:=2 \int_0^1\! d\theta\, F(\theta)\, \sin(2\pi m \theta)\, ,
\end{equation}
for $m=1,2,3,\ldots$ , with $F(\theta):=\frac{1}{2}\ln\left(\frac{\Gamma(\theta)}{\Gamma(1-\theta)}\right)-\left(\ln 2-\gamma\right)\left(\theta-1/2\right)$.

The only non-trivial integrals arising in this computation are those from the first term
in  $F(\theta)$
\begin{equation}
J_m:=\int d\theta\, \ln\left(\frac{\Gamma(\theta)}{\Gamma(1-\theta)}\right)\,\, \sin(2\pi m \theta)\,,
\end{equation}
while the second term  contributes $\left(\ln 2-\gamma\right)/(m \pi)$.

In order to proceed, we employ the expansion
\begin{equation}
\ln\left(\Gamma(\theta)\right)=-\gamma\,\theta-\ln(\theta)+\sum_{k=1}^\infty\left[\frac{\theta}{k}-\ln\left(1+\frac{\theta}{k}\right)\right]\,.
\end{equation}
With this, the integrals $J_m$ are easily calculated:
\bea
J_m\, 
&=&-2\gamma\int_0^1\!d\theta\,\theta\sin(2\pi m \theta)+
\int_0^1\!d\theta\,\sin(2\pi m \theta)\, \ln\left(\frac{1-\theta}{\theta}\right)+ \\
&&+\sum_{k=1}^\infty\int_0^1\!d\theta\,\theta\, \sin(2\pi m \theta)\, 
\left(\frac{2\,\theta+1}{k}+\ln\left(\frac{k+1-\theta}{k+\theta}\right)\right)\nonumber\\
&=&\frac{\gamma}{m\pi}+\frac{\gamma-{\rm Ci}[2\pi m]+\ln (2\pi m)}{m\pi}+\nonumber\\ &&
+\frac{1}{m\pi}\left[\sum_{k=1}^\infty\Big({\rm Ci}[2km\pi]-
{\rm Ci}[2(k+1)m\pi]\Big)-\sum_{k=1}^\infty\left(\frac{1}{k}-\ln\left(1+\frac{1}{k}\right)\right)\right],\nonumber
\eea
where ${\rm Ci}$ is the cosine integral.
Almost all terms in the sums cancel so that one eventually obtains
the simple expression
\bea
J_m&=&\frac{1}{m\pi}\left[2\gamma+\ln (2\pi m)-\lim_{N\to\infty}{\rm Ci}[2(N+1)m\pi]-\left(\ln\Gamma(1)+\gamma+\ln 1 \right)\right]\nonumber\\
&=&\frac{\gamma+\ln (2\pi m)}{m\pi}\, .
\eea
Therefore, collecting terms, we find
\begin{equation}
b_m=\frac{\gamma+\ln(2\pi m)}{m\pi}+\frac{\ln 2-\gamma}{m\pi}
=\frac{\ln(4\pi m)}{\pi m}\,,
\end{equation}
as was to be shown.

The upshot of this discussion is that we have regularised, for $0<\theta<1$, (suppressing the tadpole
and $\ln(M_s^2/\mu^2)$ terms)
\begin{equation}
- \frac{1}{\pi} \int_0^\infty\!dl\, \frac{\vartheta'_1}{\vartheta_1}(\theta,2il)
\to \frac{1}{2}\ln\left(\frac{\Gamma(\theta)}{\Gamma(1-\theta)}\right)-\left(\ln 2-\gamma\right)\left(\theta-1/2\right)\, .
\end{equation}
Now, in view of the angle condition \eqref{angle}, it is inevitable that some $\theta$s are going to be negative, so that we also have to consider the case $-1<\theta<0$. But this is easily reduced to the already derived formulas, with the result (put $-\theta=:\tilde{\theta}>0$ and apply \eqref{fourier}):
\begin{equation}
- \frac{1}{\pi} \int_0^\infty\!dl\, \frac{\vartheta'_1}{\vartheta_1}(\theta,2il)
\to \frac{1}{2}\ln\left(\frac{\Gamma(1+\theta)}{\Gamma(-\theta)}\right)-\left(\ln 2-\gamma\right)\left(\theta+1/2\right)\, ,
\end{equation}
for $-1<\theta<0$.

Now we are finally in a position to write down the complete regularised annulus amplitude \eqref{annu4ab}. The result is (still suppressing the tadpole):
\begin{multline}\label{result}
T^A({\rm D6}_a,{\rm D6}_b)
= \frac{I_{ab} N_b}{2}\Biggl[\ln\left( \frac{M_s^2}{\mu^2}\right)
\sum_{I=1}^3 \sign(\theta_{ab}^I) -\\
- \ln \prod_{I=1}^3
\left( \frac{\Gamma(|\theta_{ab}^I|)}{\Gamma(1-|\theta_{ab}^I|)} \right)^{\sign(\theta_{ab}^I)}
- \sum_{I=1}^3 \sign(\theta_{ab}^I)\,(\ln 2-\gamma)\Biggr]\,,
\end{multline}
where we have taken into account the angle condition $\eqref{angle}$.

The calculation of the M\"obius diagrams,
\bea
 T^M({\rm D6}_a,{\rm O6_k}) &=& \pm \frac{ i 4 I_{a;O6_k}}{\pi} \int_0^\infty \frac{dt}{t}\,\, \sum_{I=1}^3\,\,
         \frac{\vartheta_1'}{\vartheta_1}\left( {\textstyle i\theta_{a;O6_k}^I t,
             \frac{it}{2}+\frac{1}{2}}\right) \nonumber \\
&=& \pm \frac{4 I_{a;O6_k}}{\pi} 
 \int_0^\infty dl\,\, \sum_{I=1}^3\,\,
         \frac{\vartheta_1'}{\vartheta_1}\left({\textstyle \theta_{a;O6_k}^I,
       2il-\frac{1}{2}} \right) 
\label{moe4a},
\eea
proceeds in a rather similar fashion. Here, $I_{a;O6_k}$ is the intersection number of the D-brane and the orientifold plane $k$, $\theta_{a;O6_k}^I$ is the intersection angle of brane $a$ and the orientifold plane $k$ on the $I$'th torus and $l=1/(4t)$. The additional summand $-\frac{1}{2}$ in the second argument of the theta functions in \eqref{moe4a} leads to an additional $(-1)^{mn}=\frac{1}{2}(1+(-1)^n+(-1)^m-(-1)^{m+n})$ in the expression corresponding to \eqref{doublesumint}. Eventually, one finds
\bea
T^M({\rm D6}_a,{\rm O6_k})  &=&
 \pm I_{a;O6_k} \  \sum_{I=1}^3 \Biggl[ 4\, \theta_{a;O6_k}^I
 \left( - \ln \left(\frac{M_s^2}{\mu^2} \right)+ 2 \ln 2 - \gamma \right)  \\
&& \phantom{aaaaaaaaaa}
+ \ln \left(\frac{M_s^2}{\mu^2}\right)\, f(\theta_{a;O6_k}^I) + g(\theta_{a;O6_k}^I) \Biggr],\nonumber
\eea
where the first term vanishes after imposing the supersymmetry condition,
\bea
f(\theta) =
\begin{cases}
 \sign(\theta) \hspace{40pt} \textrm{for} -\frac{1}{2}<\theta<\frac{1}{2} \\
 \hspace{8pt} -3 \hspace{52pt} \textrm{for} -1<\theta<-\frac{1}{2} \\
 \hspace{8pt} \phantom{-}3 \hspace{52pt} \textrm{for} \phantom{-1}\frac{1}{2}<\theta<1,
\end{cases}
\eea
and
\bea
g(\theta) =
\begin{cases}
 (\gamma-3\ln2) \sign(\theta) - \sign(\theta) \ln \left(\frac{\Gamma(2|\theta|)}{\Gamma(1-2|\theta|)}\right)
                  \hspace{20pt} \textrm{for} -\frac{1}{2}<\theta<\frac{1}{2} \\
 \hspace{30pt} -3\gamma+5\ln2 + \ln \left(\frac{\Gamma(-2\theta-1)}{\Gamma(2+2\theta)}\right)
                  \hspace{56pt} \textrm{for} -1<\theta<-\frac{1}{2} \\
 \hspace{40pt} 3\gamma-5\ln2 - \ln\left( \frac{\Gamma(2\theta-1)}{\Gamma(2-2\theta)}\right)
                  \hspace{62pt} \textrm{for} \phantom{-1}\frac{1}{2}<\theta<1.
\end{cases}
\eea
The entire one loop corrections to the gauge coupling on brane stack $a$ is then given by the sum over all
annulus and M\"obius diagrams with one boundary on brane $a$.

Cases where the intersection angles sum to $\pm2n$, $n\in\mathbb{N}^*$, can be treated
by periodic continuation of our formulas (cf. \eqref{aexpr} and \eqref{fourier}).

\section{Discussion and relation to previous work}

In order to compare the derived results to \cite{Lust:2003ky}, it is useful to specialise to
$\theta_{ab}^{1,2}>0$, $\theta_{ab}^3<0$. Equation \eqref{result} then contains the following
threshold corrections
\bea
\label{thresa}
\Delta_a=-\frac{b_a}{16\pi^2}\ln\left[ \frac{\Gamma(\theta_{ab}^1) \Gamma(\theta_{ab}^2) \Gamma(1+\theta_{ab}^3)}
                {\Gamma(1-\theta_{ab}^1) \Gamma(1-\theta_{ab}^2) 
\Gamma(-\theta_{ab}^3)}\right],
\eea
($b_a=\frac{I_{ab}N_b}{2}$), which are to be compared with \cite{Lust:2003ky}
\bea
\label{thresb}
\tilde\Delta_a=-\frac{b_a}{16\pi^2}\ln\left[ \frac{\Gamma(1+\theta_{ab}^1) \Gamma(1+\theta_{ab}^2) \Gamma(1+\theta_{ab}^3)}
                {\Gamma(1-\theta_{ab}^1) \Gamma(1-\theta_{ab}^2) 
\Gamma(1-\theta_{ab}^3)}\right]\, .
\eea
Clearly, $\Delta_a$ and $\tilde\Delta_a$ are not identical,
the difference being
\bea
\label{diff}
\Delta_a-\tilde\Delta_a=-\frac{b_a}{16\pi^2}\ln\left[ -\frac{\theta_{ab}^3}{\theta_{ab}^1\theta_{ab}^2}\right]\, .
\eea
This difference appears to stem from the different treatment of open string states in the threshold corrections, which
are located at the intersection of two D6-branes and whose masses are proportional
to an integer multiple of the intersection angle $\theta_{ab}^I$. This interpretation will be motivated
in appendix \ref{comparison}. These states are
in fact included in the threshold corrections $\Delta_a$. For small intersection angles some of these states become
lighter than the string scale $M_s$, and hence $\Delta_a$ logarithmically diverges
for $\theta_{ab}^I\rightarrow 0$. 
On the other hand, $\tilde \Delta_a$ is completely regular for $\theta_{ab}^I\rightarrow 0$,
because it does not contain the contribution of these states that become light when
$\theta_{ab}^I\rightarrow0$.\footnote{Note however that
both $\Delta_a$ and $\tilde \Delta_a$ contain the contribution of states
that become light for $\theta_{ab}^I\rightarrow1$.}
In more technical terms, this different behavior can be traced back to how the infrared  divergences were
treated during the computation of the threshold corrections. In the present work,
the contribution of the massless modes appears in the
logarithmic running of the gauge coupling constant, whereas the contribution
of the  modes that become light
for $\theta_{ab}^I\rightarrow 0$ is kept in $\Delta_a$. This is in contrast to the
infrared regularisation method employed in \cite{Lust:2003ky} for the computation
of $\tilde\Delta_a$, where also the contribution of the modes with masses proportional to
$m \theta_{ab}^I$, $m\in\mathbb{N}$ is subtracted from the threshold corrections.

Finally, let us remark that the one-loop correction $\Delta_a$ to the gauge coupling
constant is not the real part of a holomorphic function when expressed in terms of the complex
structure moduli fields $U^I$ of the underlying torus $\mathbb{T}^6$, since the intersection angles
$\theta_{ab}^I$ are non-holomorphic functions of the $U^I$. The reason for this non-holomorphy
due to $\sigma$-model anomalies and other issues of holomorphy  in the context of instanton corrections
to the effective action of intersecting D-brane models are discussed  in \cite{AkerblomBlumenhagen}.


\vskip 1cm
 {\noindent  {\Large \bf Acknowledgements}}
 \vskip 0.5cm
We would like to thank Emilian Dudas, Michael Haack, Sebastian Moster, 
Erik Plauschinn, Timo Weigand and
especially Stephan Stieberger for valuable discussions.
This work is supported in part by the European Community's Human
Potential Programme under contract MRTN-CT-2004-005104
`Constituents, fundamental forces and symmetries of the universe'.


\begin{appendix}
\section{Loop channel calculation and zeta-function regularisation}\label{app}
As promised in the main text, in this appendix we determine the
analytic form of the threshold corrections using zeta-function regularisation
in the loop channel. 
The threshold corrections up to an additive constant, i.e. in particular the full
moduli dependence, will be computed.
For the annulus diagram, this means evaluating
\bea
T^A({\rm D6}_a,{\rm D6}_b)&=& \frac{i I_{ab} N_b}{2 \pi} \int_0^\infty\, \frac{dt}{t}\,\, \sum_{I=1}^3\,\,
\frac{\vartheta_1'}{\vartheta_1} \left( {\textstyle \frac{i \theta_{ab}^I t}{2},\frac{it}{2}}\right).
\eea
Using the product representation of the theta function and the Taylor expansion
$\ln(1-z)=-\sum_{n=1}^\infty \frac{z^n}{n}$
one can derive
\bea\label{appformula}
\frac{\vartheta'_1}{\vartheta_1}(\nu,\tau)=\frac{\partial}{\partial \nu} \ln \vartheta_1(\nu,\tau)
= \pi \cot(\pi \nu) - 4 \pi i \sum_{m,n=1}^\infty e^{2 \pi i \tau m n} \sinh(2 \pi i \nu n),
\eea
which is valid if  $|\exp(2 \pi i(\tau n \pm \nu))|<1$  for all  $n \in \mathbb{N}$.
Using 
\bea
\coth(x)= \sign(x)[1+2\sum_{n=1}^\infty \exp(-2 |x| m)]
\eea
and extracting the
divergence for $t \rightarrow \infty$ stemming from the massless open string modes, one finds
\bea 
\tilde\Delta&=&\int_0^\infty \frac{dt}{t} \frac{\vartheta'_1}{\vartheta_1}\left(\frac{i \theta t}{2},\frac{it}{2}\right)
+ \int_1^\infty\,  \frac{dt}{t}\, \pi i\, \sign(\theta) \label{startpointtchannel} \\ &=&
- 2 \pi i \sign(\theta) \int_0^\infty \frac{dt}{t} \sum_{n,m=1}^\infty
\Bigl[ \exp(- \pi t n(m-1+|\theta|))-\exp(-\pi t n(m-|\theta|)) \Bigr] \nonumber \\
&& - \int_0^1 \frac{dt}{t} \pi i \sign(\theta) \nonumber \\ &=&
-2 \pi i \sign(\theta) \sum_{n,m=1}^\infty \ln\left(\frac{\pi n(m-|\theta|)}{\pi n(m-1+|\theta|)}\right)
- \pi i \sign(\theta) \lim_{N\rightarrow\infty} \ln N\, . \nonumber 
\eea
Clearly, the sum over the positive integers $n$ is divergent, which was expected, as we have not
yet deducted the ultraviolet divergence due to  the tadpole.  
The main observation is that performing a simple zeta-function
regularisation $\sum_{n=1}^\infty 1 = \zeta(0) = -\frac{1}{2}$ seems to take precisely care
of the tadpole. Indeed after zeta-function regularisation we get
\bea
\tilde\Delta&=&
\pi i \sign(\theta) \sum_{m=1}^\infty
\Biggl[ \ln\left(1-\frac{|\theta|}{m}\right) - \ln\left(1-\frac{1-|\theta|}{m}\right) \Biggr]
- \pi i \sign(\theta) \lim_{N\rightarrow\infty} \ln N.\nonumber\\
&&  
\eea
Using the relations
\bea
\ln \Gamma(1+x)=-\gamma x +\sum_{k=1}^\infty\left[\frac{x}{k}-\ln(1+\frac{x}{k})\right]\ \,  {\rm and}\  
\gamma=\lim_{N\rightarrow\infty} \left( \sum_{k=1}^N \frac{1}{k} -\ln N \right)
\eea
the last expression becomes
\bea
\tilde\Delta=\pi i \sign(\theta) \ln \left( \frac{\Gamma(|\theta|)}{\Gamma(1-|\theta|)}\right)
- 2 \pi i\, \theta\, \lim_{N\rightarrow\infty} \ln N.
\eea
Finally, performing the sum over $I$ yields:
\begin{multline}
T^A({\rm D6}_a,{\rm D6}_b) =
\frac{I_{ab} N_b}{2} \sum_{I=1}^3 \sign(\theta_{ab}^I) \int_1^\infty \frac{dt}{t} \\
- \frac{I_{ab} N_b}{2} \ln \prod_{I=1}^3
\left( \frac{\Gamma(|\theta_{ab}^I|)}{\Gamma(1-|\theta_{ab}^I|)} \right)^{\sign(\theta_{ab}^I)}
+ I_{ab} N_b\,\left( \sum_{I=1}^3  \theta_{ab}^I \right)\,\, \lim_{N\rightarrow\infty} \ln (N)\, 
\end{multline}
The last term vanishes due to the supersymmetry condition \eqref{angle}. Thus, indeed, the calculation
in the loop channel using zeta-function regularisation gives the same result, up to a constant,
as the one in the tree channel after extracting the divergence that cancels due to the tadpole
condition,
 as has been done in section \ref{thresholds}. Zeta-function regularisation seems to correctly subtract
the divergence due to the closed string tadpole, an observation which we believe to be valuable as a heuristic device. Furthermore, the M\"obius diagram can be dealt with analogously if $|\theta|<\frac{1}{2}$.
\section{On light modes in the threshold corrections}
\label{comparison}
The purpose of this appendix is to provide evidence for  the statement that
the difference between the result derived here and in \cite{Lust:2003ky} is due
to the treatment of open string modes whose masses are given by an integer
multiple of an intersection angle.

Note first that if one (when working in loop channel as in Appendix \ref{app}) 
extracts not only the constant term from the (hyperbolic) cotangent, as
is done in \eqref{startpointtchannel}, but the entire cotangent, one arrives
at the result for the threshold corrections derived in \cite{Lust:2003ky}. One is thus led
to the following hypothesis: Zeta-function regularisation in the loop channel is equivalent to extracting
the cotangent from the expansion of $\frac{\vartheta'_1}{\vartheta_1}$ in \eqref{appformula}
in the tree channel (this was essentially proven in this work), whereas zeta-function regularisation in
the tree channel (which was done in \cite{Lust:2003ky}) is equivalent to extracting the cotangent
term in the loop channel\footnote{Note, when performing  the entire computation in loop-channel,
one still has to employ  zeta-function regularisation for the substraction of the 
tree-channel tadpoles.}.

Thus, the difference between the present results and the ones in \cite{Lust:2003ky} appears
to stem from the term
\bea
\int \frac{dt}{t} \pi \left[ \coth \left(\frac{\pi \theta t}{2}\right) - \sign(\theta)\right]
= 2 \pi \sign(\theta) \int \frac{dt}{t} \sum_{n=1}^\infty \exp(-\pi |\theta| t n)\,,
\label{appb1}
\eea
which can be interpreted as the contribution of modes with masses
given by $n \theta$, $n\in\mathbb{N}$.
Regularising the ultraviolet divergence in \eqref{appb1} one finds
\bea
 \int \frac{dt}{t} \sum_{n=1}^\infty
\Bigl( e^{-\pi |\theta| t n} - e^{-\pi N t n} \Bigr)
=  \zeta(0)\,  \Bigl( - \ln|\theta| + \ln N \Bigr)\,.
\eea
The finite, moduli-dependent term thus precisely accounts for the difference in \eqref{diff}.
\end{appendix}

\clearpage
\nocite{*}
\bibliography{revl}
\bibliographystyle{utphys}
\end{document}